\def\gsim{\mathop {\vtop {\ialign {##\crcr
$\hfil \displaystyle {>}\hfil $\crcr \noalign {\kern1pt \nointerlineskip }
$\,\sim$ \crcr \noalign {\kern1pt}}}}\limits}
\def\lsim{\mathop {\vtop {\ialign {##\crcr
$\hfil \displaystyle {<}\hfil $\crcr \noalign {\kern1pt \nointerlineskip }
$\,\,\sim$ \crcr \noalign {\kern1pt}}}}\limits}
\def\urusi{URu$_{2}$Si$_{2}$}

\documentstyle[seceq,preprint]{jpsj}

\def\runtitle{ Weak Antiferromagnetism
in URu$_{2}$Si$_{2}$
 }
\def\runauthor{Yukihiro {\sc Okuno} and 
Kazumasa {\sc Miyake}}

\title
{
Induced-Moment Weak Antiferromagnetism and Orbital Order \\
on the Itinerant-Localized Duality Model with Nested Fermi Surface\/: \\
A Possible Origin of Exotic Magnetism in
URu${}_{2}$Si$_{2}$
}

\author
{
Yukihiro {\sc Okuno}$^{1,2}$
\footnote{Present adress: Yukawa Institute for Theoretical Physics,
e-mail: okuno@yukawa.kyoto-u.ac.jp} and 
Kazumasa {\sc Miyake}$^{2}$
}

\inst
{
$^{1}$Department of Physics, Kyoto University, Kyoto 606-01 \\
$^{2}$Department of Physical Science,
Graduate School of Engineering Science,
Osaka University, Toyonaka 560 \\
}

\recdate
{
}
\abst
{
The weak antiferromagnetism of URu${}_{2}$Si${}_{2}$ is discussed on the
basis of a duality model which takes into account salient features of both
itinerant fermions and ``localized" component of spin degrees of freedom.
The problem is analyzed in the framework of induced-moment mechanism
by taking a singlet-singlet crystal field scheme together with the nesting
property of partial Fermi surface of itinerant fermions . It is shown that the
extremely small ordered moment $m$ of
${\cal O}$($10^{-2}$$\times$$\mu_{\rm B}$) can be
compatible with the large specific-heat jump at the transition temperature
$T_{\rm N}$.  Analysis performed in the presence of external magnetic field
shows that the field dependence of $m$ in the limit $T\to 0$ and $T_{\rm N}$
do not scale except very near the critical field $B_{\rm c}$ which
is consistent with a recent observation by Mentink {\it et al}.  It is also
shown that the antiferromagnetic magnetic order gives rise to a tiny
amount of antiferromagnetic orbital order of f-electrons.
}
\kword
{
induced-moment magnetism with nesting,
weak antiferromagnetism of URu$_{2}$Si$_{2}$, \\
itinerant-localized duality, orbital order
}
\begin{document}
\sloppy
\maketitle

\section{Introduction}
One of the major problems which have not yet been resolved in heavy Fermions
is how to understand the exotic nature of extremely weak magnetism of
\urusi\  which also exhibits a superconducting transition at the lower
temperature\cite{rf:Schlabitz}.
The most fundamental question is why the ordered moment $m$ is
so small of the order of 0.04$\mu_{\rm B}$\cite{rf:Fak,rf:Broholm1} while the
specific heat exhibits a
large jump of the order of  that in the normal state at the transition
temperature which we call $T_{\rm N}$\cite{rf:Palstra,rf:Maple}.
Moreover, it has recently been reported that the
magnetic-filed dependence of $T_{\rm N}$ and $m$ in the limit $T\to 0$ do
not seem to scale with each other but seem to have different critical fields.
\cite{rf:Mentink}  
At a glance these facts appear
to indicate that the true order parameter of this transition is not
magnetic but some hidden one which has an intimate relation with the
degrees of charge polarization\cite{rf:charge}.

However, before we look for such novel mechanisms,
it might still make sense to investigate to what extent these
anomalies can be understood by extending the conventional treatment for
magnetic mechanism so as to take into account the specific nature of \urusi .
In \urusi\ the ionic state of U$^{+4}$ with 5f$^{2}$ is
considered to be realized. Then the state with total angular momentum
$J$=4, which is formed by the Hund-rule coupling and strong spin-orbit
interaction, is split by the crystal field effect lifting the 9-fold
degeneracy into five singlets and two doublets in general.  It does not
seem unreasonable to assume that the ground state consists of 
singlet which has large matrix element of angular momentum between the
excited singlet state. 
For instance, the singlet-singlet model
was adopted to explain the temperature dependence of the bulk magnetic
susceptibility and the heat capacity,~\cite{rf:Nieuwenhuys,rf:Radwanski}
and the collective excitations measured by the inelastic neutron
scattering\cite{rf:Broholm2}.

Although the singlet ground state has no average
magnetic moment, the so-called induced-moment mechanism can work 
to cause the ordered state 
by invoking the virtual process of mixing between the ground state and
the low-lying excited states of crystal field.  The ordered moment so
obtained is rather reduced below its full moment in general
\cite{rf:Wang,rf:Grover}.  However, the usual induced-moment mechanism
on the localized model is not enough to understand simultaneously the
extremely small magnetic moment of the order of 1\% of the full uranium
moment and the large specific heat jump at $T_{\rm N}$.

In this paper we demonstrate that the fundamental property of ``magnetic
transition" of \urusi\ can be understood by the induced-moment mechanism
on the basis of a itinerant-localized duality model together with an
assumption of the nesting of the part of Fermi surface.
Sikkema {\it et al} have already proposed the model 
with the nested Fermi surface
on the  singlet-singlet crystal field scheme
for  the weak-moment formation of
URu${}_{2}$Si${}_{2}$~\cite{rf:Sikkema}. Their model, an Ising-Kondo lattice
model with transverse field which originates from the off-diagonal element
of two singlet levels, produces weak moment but rather small transition
temperature and does not reproduce the large specific heat jump.
Our scheme here shows  that
the tiny moment of antiferomagnetic order does not contradict with large
specific heat jump at $T=T_{\rm N}$, and 
the unusual magnetic-field dependence of
$m(T\to 0)$ and $T_{\rm N}$ can be reproduced within the order of magnitude
for a reasonable set of parameters.  
It is also shown that the charge distributions of f-electrons at
two sublattice sites in the antiferromagnetic state are different
giving rise to a tiny amount of antiferromagnetic orbital order.
\\

\section{Formalism}
\subsection {Outline of itinerant-localized duality model}
An itinerant-localized duality model has been proposed by
Kuramoto and Miyake~\cite{rf:Kuramoto1,rf:Kuramoto2} as a
quantum phenomenology in order to explain the properties of heavy Fermions
beyond the conventional Fermi liquid such as the weak antiferromagnetism and
metamagnetism.
Quite recently, an investigation to examine its microscopic basis was put
forth.~\cite{rf:Okuno}  The problem about exotic magnetism of
URu${}_{2}$Si${}_{2}$ seems to within the scope which the duality model
can be applied as was already argued in its simple fasion.~\cite{rf:Kuramoto2}

In the ``duality model", the partition function ${\rm Z}$ and the
effective action ${\rm A}$ is given as~\cite{rf:Kuramoto1,rf:Kuramoto2}
\begin{subeqnarray}
Z&=&\int{\rm D}f^{\dagger}\int{\rm D}f\int{\rm D}{\bf S}
{\rm exp}(-\beta A) \label{2.1a} \\
{\it A}&=&A_{{\rm f}} + A_{{\rm s}} +A_{{\rm int}},
\label{2.1b}
\end{subeqnarray}
where $A$'s are defined in terms of $f$ and  $f^{\dagger}$, Grassmann
numbers, and ${\bf S}$, {\it c}-number vector, as follows:
\begin{subeqnarray}
A_{{\rm f}}&=&
-\sum_{i,j.\sigma}\sum_{n}f^{\dagger}_{i \sigma}(-{\rm i}\epsilon_{n})
(G^{-1}_{ij,\sigma}
({\rm i}\epsilon_{n}))f_{j \sigma}({\rm i}\epsilon_{n}),
\label{2.2a} \\
A_{{\rm s}}&=&
\frac{1}{2}\sum_{i,j,m}{\bf S}{}_{i}(-{\rm i} \nu_{m})
(\chi_{0 ij}^{-1}(i \nu_{m})\delta_{ij}-J_{ij}){\bf S}{}_{j}({\rm i} \nu_{m})
-\sum_{i}h_{i}S_{iz}
\label{2.2b}\\
A_{{\rm int}}&=&
-\lambda_{0}\sum_{i \alpha \beta}\sum_{mn}f^{\dagger}_{i \alpha}
(-{\rm i}\epsilon_{n}-{\rm i}\nu_{m})
f_{i \beta}({\rm i}\epsilon_{n}){\vec{\sigma}}_{\alpha \beta}
\cdot{\bf S}{}_{i}({\rm i}\nu_{m}),
\label{2.2c}
\end{subeqnarray}
where $A_{{\rm f}}$ represents the part of itinerant fermions,
$A_{{\rm s}}$ the localized component of spin degrees of freedom
consisting of incoherent part of fermions,
and $A_{{\rm int}}$ is the interaction between the fermion  and
the ``localized spin".  $G({\rm i}\epsilon_{n})$ in $A_{\rm f}$,
eq.(\ref{2.2a}), is the Green
function of the itinerant fermion and $\chi_{0}({\rm i}\nu_{m})$ in
$A_{\rm s}$, eq.(\ref{2.2b}), is the partially renormalized local spin
susceptibility which does not include the effect of neither the RKKY
interaction and nor the coupling with fermions.  The exchange interaction
between  ``localized spins" is represented by $J_{ij}$, and
a magnetic field at site $i$ is denoted by $h_{i}$. 
Although the above form of the action is isotropic 
in spin space we retain only one component, $S_{z}$, because 
the large uniaxial magnetic anisotropy exists in the case we 
discuss below. 
So in the discussion below we only retain one component of spin, $S_{z}$. \\
\indent In order to discuss the magnetic 
properties of the strongly correlated
systems, we first take the trace over $f^{\dagger}$ and $f$ as follows:
\begin{subeqnarray}
Z&=&{\rm det}G\times \int {\rm D}S {\rm exp}(-\beta A_{\rm m})
\label{2.3a}\\
A_{\rm m}&=&A_{\rm s}-{1\over \beta}{\rm Tr}{\rm ln}(1+\lambda_{0} G
\sigma_{z}\cdot S_{z}).
\label{2.3b}
\end{subeqnarray}
As a first step of approximation, we take the saddle-point approximation
for the macroscopic mode ${\bf S}_{q}$ (with the wavevector $q=0$ or
the antifferomaginetic wavevector $Q$) and the
Gaussian average with respect to other modes of spin fluctuations.  Then,
we obtain the equations of states in the form
\begin{subeqnarray}
\biggl[\frac{1}{\chi_{0}(0,S)}-J(0)
-2\lambda_{0}^{2}\Pi (0,S_{0})\biggr] S_{0}&=&h \label{det1} \\
\frac{1}{\chi_{0}(Q,S)}-J(Q)
-2\lambda_{0}^{2}\Pi (Q,S_{Q})&=&0, \label{det2}
\end{subeqnarray}
where $J(q)$ (with $q=0$ or $Q$) is the Fourier component of the exchange
interaction $J_{ij}$, and $\chi_{0}(q,{\bf S})$ is the Fourier component
of the static ``local susceptibility" which has site dependence in general.
When we take into account only
the nearest-neighbor interaction, $J(0)=-zJ$ and $J(Q)=zJ$,
$z$ being the number of nearest neighbors.  It is noted that the local
susceptibility $\chi_{0ii}$ has a site dependence due to the mean field of
antiferromagnetic induced moment, in general, as discussed below.
As a result, $\chi_{0}$ has a wavenumber dependence as in eqs.(\ref{det1}).
The polarization function $\Pi(q,{\bf S}_{q})$ of the itinerant component is
given as
\begin{eqnarray}
\Pi (q,{\bf S}_{q}) = -\frac{1}{N}\sum_{{\bf k}}\sum_{\epsilon_{n}}
[G_{\sigma}({\bf k},{\rm i}\epsilon_{n})^{-1}
G_{\sigma}({\bf k}+{\bf q},{\rm i}\epsilon_{n})^{-1}
-(\lambda {\bf S}_{q})^{2}]^{-1}.
\label{polarization}
\end{eqnarray}
\\

\subsection{Induced-moment antiferromagnetism}
A structure of eq.(\ref{det2}b) is the same as the mean field equation in
the induced-moment mechanism except for the effective exchange interaction
due to the polarization $\Pi(q,S_{Q})$.  
In the latter mechanism, $\chi_{0}$ is given
by the so-called Van Vleck susceptibility arising from the virtual
excitation between the singlet ground state and some excited
state of crystal field levels\cite{rf:Wang,rf:Grover}.
In the present problem, $\chi_{0}$ is
considered to consist both of such local Van Vleck susceptibility and
Kondo like correlation arising from the effect of hybridization between
the localized f-electron and the conduction electrons in general.
In order to obtain the extremely small ordered moment
$m\sim{\cal O}(10^{-2}$$\times$$\mu_{\rm B}$), the conventional induced-moment
mechanism is not enough unless the exchange interaction $J(Q)$ almost
coincides accidentally with its threshold for the occurrence of ordered state
\cite{rf:Grover}.
Furthermore, only a tiny jump in the specific heat can be expected in such
a situation.  So, the
effect of polarization should be crucial as pointed out in ref.15.
We evaluate $\Pi(q,{\bf S}_{q})$, eq.(\ref{polarization}), with use of the
quasi-particle form for $G_{\sigma}(k,{\rm i}\epsilon_{n})$:
\begin{equation}
G_{\sigma}(k,{\rm i}\epsilon_{n})=
{a_{\rm f}\over{\rm i}\epsilon_{n}-\sigma \tilde{h}-E_{k}},
\label{green}
\end{equation}
where $a_{\rm f}$ is the renormalization factor, $E_{k}$ is
the energy spectrum of the itinerant fermion measured from the
chemical potential, and $\tilde{h}=a_{\rm f}h$ is
the renormalized external field.  It is noted that the renormalized
field $\tilde{h}$ is very small value $(\tilde{h} << h)$ in the strongly
correlated regime, while the ``localized spin'' is affected directly
by the external field $h$ which affects the itinerant fermion through
the coupling $\lambda_{0}$ in, 
 $A_{\rm int}$, eq.(\ref{2.2c}).  As for the form of
$\chi_{0}$, the local spin susceptibility, we take so as to reproduce
the Van Vleck susceptibility on the singlet-singlet scheme neglecting
the effect of Kondo like correlation.
As the suitable form of the local susceptibility, 
$\chi_{0}(0,S)$ and $\chi_{0}(Q,S)$, we take those form  
in the case of induced moment as explained below. \\
\indent Assuming the two sublattices $A$ and $B$, the molecular-field
equation of induced-moment antiferromagnetism is
given as follow\cite{rf:Grover}:
\begin{subeqnarray}
S_{A}&=&\frac{-2(zJS_{B}-h){\rm c}^{2}}
{\sqrt{\Delta^{2}+(2zJS_{B}-2h)^{2}{\rm c}^{2}}}
{\rm th}\frac{\sqrt{\Delta^{2}+(2zJS_{B}-2h)^{2}{\rm c}^{2}}}{2T}
\label{sa}\\
S_{B}&=&\frac{-2(zJS_{A}-h){\rm c}^{2}}
{\sqrt{\Delta^{2}+(2zJS_{A}-2h)^{2}{\rm c}^{2}}}
{\rm th}\frac{\sqrt{\Delta^{2}+(2zJS_{A}-2h)^{2}{\rm c}^{2}}}{2T},
\label{sb}
\end{subeqnarray}
where $S_{A}$ and $S_{B}$ are the induced moment at the site $A$ and $B$,
respectively.  $\Delta$ denotes the energy difference
between the ground-state singlet, $\vert 0\rangle$, and the excited-state
singlet, $\vert 1\rangle$,
and ${\rm c}\equiv\langle 1 \vert {\rm J}_{z}\vert 0 \rangle$, which is an
essential ingredient of the Van Vleck susceptibility.
In deriving eqs.(\ref{sa}a) and (\ref{sb}b), we have noted that the effective
field at the site A is given by $h_{A  {\rm eff}}=h-zJS_{B}$, and that at
the site B is $h_{B  {\rm eff}}=h-zJS_{A}$, respectively.

The local spin susceptibility $\chi_{0}(0,S)$'s in
eqs.(\ref{det1}a) and (\ref{det2}b) are defined from the
relations as follows:
\begin{subeqnarray}
S_{0} &=& h_{0}\chi_{0}(0,S)
\label{meanfield1} \\
S_{Q} &=& h_{Q}\chi_{0}(Q,S),
\label{meanfield2}
\end{subeqnarray}
where $h_{q}$ and $S_{q}$, with $q=0$ or $Q$,
are the Fourier component of the effective
field and the induced moment, respectively, defined as
\begin{subeqnarray}
h_{0} &=& \frac{1}{N}\sum_{r_{i}}{\rm e}^{{\rm i}0\cdot r_{i}}
h_{{\rm eff},i}
=\frac{1}{2}(h_{{\rm eff},A}+h_{{\rm eff},B})
=h-\frac{1}{2}(zJS_{A}+zJS_{B})
\label{2.9a} \\
h_{Q} &=& \frac{1}{N}\sum_{r_{i}}{\rm e}^{{\rm i}Q\cdot r_{i}}
h_{{\rm eff},i}
=\frac{1}{2}(h_{{\rm eff},A}-h_{{\rm eff},B})
=\frac{1}{2}(zJS_{A}-zJS_{B})
\label{2.9b} \\
S_{0} &=& \frac{1}{N}\sum_{r_{i}}{\rm e}^{{\rm i}0\cdot r_{i}}S_{i}
=\frac{1}{2}(S_{A}+S_{B})
\label{2.9c} \\
S_{Q} &=& \frac{1}{N}\sum_{r_{i}}{\rm e}^{{\rm i}Q\cdot r_{i}}S_{i}
=\frac{1}{2}(S_{A}-S_{B}).
\label{2.9d}
\end{subeqnarray}
Here susceptibility defined above contains the non-linear components
of effective field in general.  
However, when the effective field is small as in the 
present case, we can
approximate it as the linearized susceptibility.
Then, substituting (\ref{2.9a}a) [(\ref{2.9b}b)] and (\ref{2.9c}c)
[(\ref{2.9d}d)] with (\ref{sa}a) and (\ref{sb}b), we obtain the expression
for the susceptibility $\chi_{0}(0,S)$ [$\chi_{0}(Q,S)$]
as follows:
\begin{subeqnarray}
\chi_{0}(0,S)&=&\frac{\frac{1}{2}(h-zJS_{B})\chi_{A}
+\frac{1}{2}(h-zJS_{A})\chi_{B}}{h-\frac{1}{2}(zJS_{A}+zJS_{B})}
\label{chi1} \\
\chi_{0}(Q,S) &=& \frac{\frac{1}{2}(h-zJS_{B})\chi_{A}
-\frac{1}{2}(h-zJS_{A})\chi_{B}}{\frac{1}{2}(zJS_{A}-zJS_{B})},
\label{chi2}
\end{subeqnarray}
where  $\chi_{A}$ and $\chi_{B}$ are defined as
\begin{subeqnarray}
\chi_{A} &=& \frac{2{\rm c}^{2}}
{\sqrt{\Delta^{2}+(2zJS_{B}-2h)^{2}{\rm c}^{2}}}
{\rm th}\frac{\sqrt{\Delta^{2}+(2zJS_{B}-2h)^{2}{\rm c}^{2}}}{2T}
\label{chia} \\
\chi_{B} &=& \frac{2{\rm c}^{2}}
{\sqrt{\Delta^{2}+(2zJS_{A}-2h)^{2}{\rm c}^{2}}}
{\rm th}\frac{\sqrt{\Delta^{2}+(2zJS_{A}-2h)^{2}{\rm c}^{2}}}{2T}.\label{chiAB}
\label{chib}
\end{subeqnarray}
\\

\section{Analysis of the Model}
First of all, let us assess the size of parameters appearing in  the above
formulae keeping it in mind to applying them to  \urusi.
The crystal field parameter $\Delta$ 
and the matrix element $c$ can be estimated so as to reproduce the
temperature dependence of susceptibility in high temperature region.
In Niewenhuys's scheme of crystal field~\cite{rf:Nieuwenhuys}, 
$\Delta$=1.2$\times10^{2}$ K and $c$=1.2$\mu_{{\rm B}}$, the values of 
which have been frequently used by analysis. 
On the other hand, in the scheme 
of Santini {\it et al}'s,~\cite{rf:charge} 
$\Delta$=4.6$\times10^{2}$ K and $c=1.6\mu_{{\rm B}}$, the values of which 
give better agreement with experiment.
Therefore, we adopt the latter scheme by Santini {\it et al}. 
In the latter  crystal field scheme, there exist doublet 
levels between the ground and the excited 
singlet levels, and this fact explains naturally the large entropy
at low temperature leading to the large specific heat.   
The exchange interaction $J$ is chosen so as to fit
the neutron scattering data on the basis of RPA theory of singlet-singlet
model.~\cite{rf:Broholm2}  
The `band' width of itinerant fermion, $2D$, is of the order of
$T_{\rm coh}$ which is about 100K for URu${}_{2}$Si${}_{2}$.
The renormalized spin-fermion coupling $\tilde{\lambda}$
is the same order of $T_{\rm coh}=T_{\rm K}$~\cite{rf:Kuramoto1}.
In the calculations below,
we set the matrix element ${\rm c}$ and $D$ as unit of magnetization
and of energy, respectively. Here we remark about the 
values of parameters. Since various assumptions are involved in the
above estimation, there remains ambiguity about these values. 
In particular  
the matrix element $c$ is difficult to evaluate and actual value 
 may be possibly smaller than the estimated one.  
So is  the crystal field splitting $\Delta$, and it can be much larger.
We discuss these points later.  
\\

\subsection{Ordered moment at zero temperature}
When the external field is absent, $h=0$, the uniform component of magnetic
moment $S_{0}=0$.  Then, by substituting (\ref{green}) into
(\ref{polarization}), the polarization $\Pi$ in eq.(\ref{det2}b)
is expressed as
\begin{equation}
\Pi (Q,S_{Q})=-\frac{a_{\rm f}^{2}}{N}T\sum_{\epsilon_{n}}\sum_{{\bf k}}
[({\rm i}\epsilon_{n}-E_{\bf k})({\rm i}\epsilon_{n}-E_{{\bf k}+{\bf Q}})
-({\tilde{\lambda}}S_{Q})^{2}]^{-1},
\label{piq1}
\end{equation}
where $\tilde{\lambda}\equiv a_{\rm f}\lambda_{0}$.  If the nesting condition
$E_{{\bf k}+{\bf Q}}=-E_{{\bf k}}$ is fulfilled, (\ref{piq1}) is easily
evaluated as
\begin{eqnarray}
\Pi (Q,S_{Q}) &=& {1\over 2}a_{\rm f}^{2}\rho_{\rm F}
\int_{-D}^{D}{{\rm d}\epsilon\over\sqrt{\epsilon^{2}+({\tilde{\lambda}}S_{Q})^{2}}}
{\rm th}{\sqrt{\epsilon^{2}+({\tilde{\lambda}}S_{Q})^{2}}\over 2T},
\label{piq2}
\end{eqnarray}
where $\rho_{\rm F}$ is the density of states at the Fermi level and
$D$ is half the bandwidth of itinerant fermion.

Hereafter, we use $m$ for the staggered magnetization, i.e., $S_{Q}=m$.
At $T=0$, $\Pi(0,m)$ is estimated as
\begin{eqnarray}
\Pi (Q,m) &=& a_{\rm f}^{2}\rho_{\rm F}
{\rm log}\frac{2D}{\tilde{\lambda}m}.
\label{piq3}
\end{eqnarray}
The local susceptibility $\chi_{0}$ in eq.(\ref{det2}b)
at $T=0$ is evaluated from (\ref{chib}b) as
\begin{subeqnarray}
\chi_{0} (0,S_{Q}) &=& \frac{2{\rm c}^{2}}
{\sqrt{\Delta^{2}+(2zJ{\rm c}m)^{2}}} \\
& \simeq & \frac{2{\rm c}^{2}}{\Delta}. \label{chi0}
\end{subeqnarray}
The last approximation in (\ref{chi0}) holds when the level splitting
$\Delta$ of crystal field is far larger than $2zJcm$.

Thus the equation of state (\ref{det2}b) at $T=0$ is reduced to
\begin{eqnarray}
{\sqrt{\Delta^{2}+(2zJ{\rm c}m)^{2}}\over 2{\rm c}^{2}}
-zJ-2\tilde{\lambda}^{2}\rho_{\rm F}
{\rm log}\frac{2D}{\tilde{\lambda}m} &=& 0.
\label{eqofstate}
\end{eqnarray}
If the approximation of (\ref{chi0}b) is valid, we can estimate the size of
the magnetization $m$ at $T=0$ as
\begin{equation}
m \simeq \frac{2D}{\tilde{\lambda}}
{\rm exp}\bigl[-\frac{1}{2\tilde{\lambda}^{2}
\rho_{\rm F}}(\frac{\Delta}{2{\rm c}^{2}}-zJ)\bigr], \label{SDW} \\
\end{equation}
which is meaningful only in the case $\Delta>2zJc^{2}$, where the ordered
state is not realized without a help of the polarization $\Pi$ because
the conventional condition for the induced-moment ordering to occur
is given by $\Delta/2{\rm c}^{2}<zJ$.
The ordered moment given by (\ref{SDW}) can become extremely small only
for the small value of the local susceptibility, i.e.,
$\chi_{0}(Q,S)/\rho_{\rm F}\simeq 2c^{2}/\Delta\rho_{\rm F}\ll 1$
because ${\tilde{\lambda}}\rho_{\rm F}\sim 1$ and $D/{\tilde{\lambda}}\sim 1$
as mentioned above.  It is noted that such an extremely small ordered moment
is hard to be realized only from the usual nesting property of
itinerant fermions unless $T_{\rm N}\ll D$ which is not the case in the
present problem.  While the ordered moment is extremely small, the specific
heat exhibits a rather large jump at $T=T_{\rm N}$ of the order of
$C_{\rm n}(T_{\rm N})$, the specific heat at the normal side, as will
be discussed below.  This is because
the mathematical structure of the thermodynamic potential in the
nesting system is similar to that of the superconductivity.
\\

\subsection{Temperature dependence of ordered moment}
The temperature dependence of $m$ is obtained from the relation
(\ref{det2}b)
\begin{equation}
\frac{\sqrt{\Delta^{2}+(2zJm{\rm c})^{2}}}{2{\rm c}^{2}}
{\rm cth}\frac{\sqrt{\Delta^{2}+(2zJm{\rm c})^{2}}}{2T}-zJ
-{\tilde{\lambda}}^{2}
\rho_{\rm F}
\int_{-D}^{D}{{\rm d}\epsilon\over\sqrt{\epsilon^{2}+({\tilde{\lambda}}m)^{2}}}
{\rm th}{\sqrt{\epsilon^{2}+({\tilde{\lambda}}m)^{2}}\over 2T} = 0,
\label{eqofstatet}
\end{equation}
where we have used (\ref{piq2}) for $\Pi(Q,m)$ and (\ref{chi2}b) for
$\chi_{0}(0,m)$ with (\ref{chia}) substituted by $h=0$.
When the transition temperature is small compared to  $\Delta$
we can neglect the $m$ dependence of the localized-spin part of
susceptibility and can estimate the transition temperature as,
\begin{equation}
T_{N} \simeq 2.26D
{\rm exp}\bigl[-\frac{1}{2\tilde{\lambda}^{2}
\rho_{\rm F}}(\frac{\Delta}{2{\rm c}^{2}}-zJ)\bigr]. \label{Tnapp} \\
\end{equation}
So we can get the relation $T_{N}\simeq {1.13\tilde{\lambda}}m$ from
(\ref{SDW})~\cite{rf:Kuramoto2}. Considering that the $\tilde{\lambda}$ 
is  the order of $T_{coh}$, we expect relatively large value of transition 
temperature compare to the magnitude of the moment from this scaling 
relation. \\
\indent Numerical solutions of (\ref{eqofstatet}) are shown in Fig.\ 1 where
the temperature dependence of the magnetization $m$ for various
values of $\Delta$ and $\tilde{\lambda}$ are drawn.  
The realization of small value of magnetization stands in the
delicate balance of various parameters which is considered to be realized
in URu${}_{2}$Si${}_{2}$.
One peculiar feature of URuSi$_{2}$ is the extremely small value of
magnetization compared to rather `high' transition temperature $T_{N}$.
 In order to obtain the  extremely small magnetization $m$ but 
 not so small value of $T_{N}$,
 the equations (\ref{SDW}) and (\ref{Tnapp})
 suggest that it requires not
 only large value of spin-fermion coupling $\tilde{\lambda}$ but
also the large value of crystal field splitting.
Recalling that we have taken $c$ and $D$ as the unit of the value of
magnetization and energy, respectively, the dimension of our parameter
is $[J]=[E]/[M]^{2}$, $[\tilde{\lambda}]=[E]/[M]$, $[h]=[E]/[M]$, 
$[\rho_{F}]=[E]{}^{-1}$,
$\Delta=[E]$ and $[T]=[E]$, where the [E] is the unit of energy and 
[M] is that of magnetization. Therefore, when we make a correspondence 
between the actual value of magnetization and our calculated
 $m$, and the actual transition
temperature and our calculated $T_{N}$, we must multiply $m$ by the real 
value of $c$ for magnetization and $D$ for transition temperature. 
The value of $\rho_{F}$ are determined in the range where it is consistent 
to the specific heat jump observed by the experiment. 
For example, with setting the energy unit as $D$=100K, 
the value $\rho_{F}=0.2$ gives the linear 
specific heat coefficient 
$\gamma \equiv 2\pi^{2}k_{{\rm B}}^{2}\rho_{F}/3$ as 
$110{\rm mJ}/{\rm K^{2}}{\rm mol}$ 
which is about twince the decrease of $\gamma$, 
$50{\rm mJ}/{\rm K^{2}}{\rm mol}$, across 
the magnetic trantision.\cite{rf:Maple} \\
\indent The smallness of magnetization may be a manifestation 
of the fact that the actual 
value of $\Delta$ is larger than that frequently used so far.
It is rather difficult task
to determine the actual value of $c$, $D$ and $\Delta$
from experiments so that their frequently used  value
($c=1.2\mu_{{\rm B}}$, $D\sim 100K$ and $\Delta =120K$) 
have ambiguity to some extent. 
In particular, if we estimate the Van Vleck contribution to 
the magnetic susceptibility, $2c^{2}/\Delta$, with
$\Delta=120K$ and
$c=1.2\mu_{{\rm B}}$, we obtain twice as large value as the observed 
 magnetic susceptibility ($\sim 5*10^{-3}$emu/mol)\cite{rf:Fak}
at the zero temperature. Considering the fact that the observed value 
of the susceptibility further contains the contribution from the 
itinerant quasiparticles in general, we must take larger 
value of $\Delta$  than $\Delta=120K$. 
In fact Santini {\it et al} have taken $\Delta=460K$ and $c=1.6\mu_{{\rm B}}$ 
to reproduce the obtained the susceptibility within the singlet 
ground crystal field scheme\cite{rf:charge}.
So it is reasonable to take the large value of $\Delta$ as 
in our calculation. If we take $D=100K$ and $c=1.6\mu_{{\rm B}}$, our result
of solid line in  Fig.\ 1 
implies that $T_{N}$ is about 7K and $m$ is about 0.13$\mu_{{\rm B}}$. 
The magnitude of these values  
 are different from those of observed values ($T_{N}$=17.5K and 
$m$=0.04$\mu_{{\rm B}}$) about numerical factors, but  
we obtain the  same order as those of observed values of $T_{N}$ and $m$. 
The difference in the factor may be improved by changing the  
parameter value within the permitted range. 
 For example, if we take much larger value for $\tilde{\lambda}$ 
and little value for $\rho_{F}$, we will obtain better result. 
 Another improvement will be 
achieved  if we consider the unit of energy $D$ to be larger 
than ($D=$)100K.
$D$ is the `band width' 
of itinerant fermion in the duality model and  is further renormalized 
to the Fermi liquid fixed point. So the value of $D$ has ambiguity of 
order $1$. If we set $D$=200K, for example, our result of $T_{N}$
reproduces the experimental value without making the other
parameters  out of the permitted range.  
Finally we should not forget the possibility of  more suitable 
scheme of crystal field levels. We are based on the scheme of 
Santini {\it el al}'s. However the difficulties of determination of  
crystal field levels remain the room for much larger 
(smaller) value of $\Delta$ ($c$) so that we 
might obtain better result for the $T_{N}$ and $m$.       
\begin{figure}
\vspace*{10cm}
\caption{Temperature dependence of the ordered moment 
for various values of parameters. The parameters here
are normalized by the unit $c$ for the magnetization 
and $D$ for the energy.}
\label{fig:1}
\end{figure}

\subsection{Specific heat jump}
The specific heat jump $\Delta C$ at $T=T_{\rm N}$ is given in terms of the
coefficients $\alpha$ and $\beta$ of the Ginzburg-Landau free energy as
\begin{equation}
\Delta C=T_{\rm N}\frac{\alpha^{2}}{\beta}.
\label{jump}
\end{equation}
Without the $m$-dependence of $\chi_{0}$, these coefficients are the same
as those of BCS.  The deviations from BCS are determined by
expanding the equation of state, (\ref{det2}b), with respect to $m$, and
comparing the linear and cubic terms as follows:
\begin{subeqnarray}
\alpha &=& \bigl(\tilde{\lambda}^{2}\rho_{{\rm F}}
+\frac{1}{2{\rm sinh}^{2}\frac{\Delta^{2}}{2T_{\rm N}}}
\frac{\Delta}{8c^{2}T_{\rm N}}\bigr)
\frac{1}{T_{\rm N}}
\label{alpha}
\\
\beta &=& \rho_{{\rm F}}\tilde{\lambda}^{4}\frac{7\zeta (3)}{8\pi^{2}
T_{\rm N}^{2}}+
\frac{(zJ)^{2}}{2\Delta}{\rm cth}\frac{\Delta}{2T_{\rm N}}
-\frac{(zJ)^{2}}{2T_{\rm N}}\frac{1}{{\rm sinh}^{2}
\frac{\Delta}{2T_{\rm N}}}
\label{beta}
\end{subeqnarray}
where the first terms are those for BCS.
In the case of URu${}_{2}$Si${}_{2}$, it is expected that
$\rho_{{\rm F}}\tilde{\lambda}\sim 1$, $\Delta\sim\tilde{\lambda}
\gsim zJ$, and $\Delta$ is several times larger than $T_{\rm N}$; so that
the  first terms in (\ref{alpha}a) and (\ref{beta}b) are predominant
reproducing the BCS value for $\Delta C$.  Namely, we obtain
a large specific heat jump as in the BCS or simple SDW systems and 
actually we reproduce the same order of experimental value of 
specific heat jump with the typical values of our parameters in 
section 3.3.
The reason why Sikkema {\it et al} cannot reproduce the large 
specific heat jump~\cite{rf:Sikkema} is mainly due to their value of 
$\Delta$. They have taken the value 120K for $\Delta$ by seeing the 
balance of other parameters in their simple Hamiltonian.       
We take rather large value of $\Delta$ and 
does not attach importance to keep the $\Delta$=120K because 
 its value estimated from the experiment has ambiguity to some extent.
Furthermore their specific heat jump are mainly due to the 
 gap opening of the Fermi surface of the 
conduction electron which are not renormalized, 
but within our scheme the itinerant electron are renormalized 
and they bring large specific heat jump. \\

\subsection{Orbital order induced by antiferromagnetism}
The induced mean-field $h_{\rm eff}$ determines the orbital state of
f-electrons through the matrix element $c=<1|J_{z}|0>$.  Indeed, the effective
Hamiltonian $H^{\rm loc}_{\rm eff}$ for the local state is written as
\begin{equation}
H^{\rm loc}_{\rm eff}=E_{0}|0><0|+(E_{0}+\Delta)|1><1|
-h_{\rm eff}c\bigl(|0><1|+|1><0|\bigr),
\label{orbit}
\end{equation}
where $E_{0}$ is the energy of the crystal field state $|0>$.  The ground
state $|{\rm g}>$ of (\ref{orbit}) is given by
\begin{equation}
|{\rm g}>={1\over
\sqrt{(h_{\rm eff}c)^{2}+(E_{\rm g}-E_{0})^{2}}}\bigl(h_{\rm eff}c|0>
+(E_{\rm g}-E_{0})|1>\bigr),
\label{groundstate}
\end{equation}
where the ground state energy $E_{\rm g}$ is defined as
\begin{equation}
E_{\rm g}=E_{0}+{1\over 2}\bigl[\Delta-\sqrt{\Delta^{2}+4(h_{\rm eff}c)^{2}}
\bigr]
\label{groundenergy}
\end{equation}
This means that the local orbital state depends on the mean field
$h_{\rm eff}$.  Therefore, the antiferromagnetic induced-moment, i.e.,
the mean field, gives rise to the  orbital order of f-electrons
because the induced moment is different between the sublattices.  Note that
the charge density corresponding to the state (\ref{groundstate})
depends on the sign of $h_{\rm eff}$ through the cross term of $|0>$ and
$|1>$.  This may be a novel aspect of induced-moment antiferromagnetism which
has not yet been recognized so far. This effect will give some physical 
response that suggests at first sight 
some `hidden order' other than the antifferomagnetic 
order. But when induced moment is small as in this case, 
it is hard to detect the orbital order. Indeed ${}^{101}$Ru NQR 
study has not detected the 
explicit signal of orbital order in URu${}_{2}$Si${}_{2}$.~\cite{rf:Kohori} 
\\

\subsection{Effects of external field}
Here let us discuss the effect of external field on
the ordered moment $m$ at zero temperature and the transition
temperature $T_{\rm N}$.  First, $m=\frac{1}{2}(S_{A}-S_{B})$ is
determined from eqs.(\ref{det1}) as follows:
\begin{subeqnarray}
\frac{1}{\chi_{0}(Q,S)} -zJ-2\tilde{\lambda}^{2}\rho_{\rm F}
\Pi (Q,m;h)&=& 0
\label{sqh} \\
\bigl(\frac{1}{\chi_{0}(0,S)}+zJ
-2\tilde{\lambda^{2}}\rho_{\rm F}\bigr)S_{0} &=& h. \label{magdet}
\end{subeqnarray}
where the polarization $\Pi$ is given as
\begin{subeqnarray}
\Pi (Q,m;h) &=&
{\rm log}\frac{D+\sqrt{D^{2}+(\tilde{\lambda} m)^{2}}}
{\tilde{h}+\sqrt{\tilde{h}^{2}-(\tilde{\lambda} m)^{2}}}
 \hspace{0.5cm} (\tilde{h}>\tilde{\lambda} m) \\
 &=& {\rm log}
\frac{D+\sqrt{D^{2}+(\tilde{\lambda} m)^{2}}}
{\tilde{\lambda} m}
\hspace{0.5cm} (\tilde{\lambda} m>\tilde{h}).
\end{subeqnarray}

Next, $T_{\rm N}$ under the external field is determined by the relations
similar to eqs.(\ref{sqh}):
\begin{subeqnarray}
\frac{1}{\chi_{0}(Q,S)}-zJ-\frac{\tilde{\lambda}^{2}\rho_{\rm F}}
{2}\int_{-D}^{D}\frac{{\rm d}\epsilon}{\epsilon}
\bigl({\rm th}\frac{\epsilon+\tilde{h}}{2T{}_{N}}+
{\rm th}\frac{\epsilon-\tilde{h}}{2T{}_{N}}\bigr) &=& 0 \\
\frac{S_{0}}{\chi_{0}(0,S)}+zJS_{0}
-2\tilde{\lambda}^{2}\rho_{\rm F}S_{0} &=& h. \label{Tndet}
\end{subeqnarray}

The critical field $h_{\rm c}$ given by (\ref{magdet}), where $m\to 0$,
and by (\ref{Tndet}), where $T_{N}\rightarrow 0$, coincides with each other
of course.  However, the $h$-dependence of $m$ and $T_{\rm N}$ at
lower field $h<h_{\rm c}$ can exhibit rather different behaviors.
Indeed, the numerical solutions of (\ref{magdet}) and
(\ref{Tndet}) are shown in Fig.\ 2 for the parameters seemingly relevant to
\urusi.  In Fig.\ 2 the numerical instability occurs for high field
region $h\simeq h_{\rm c}$ in determining $m$ in (\ref{magdet}),
while we are concerned with the lower field region.
We can see in Fig.\ 2 that the reduction rate of $m$ as increasing $h$
is much larger than that of $T_{\rm N}$ in the low field region.
This result can be understood qualitatively as follows.  The degree
of nesting is rapidly destroyed by the external field at $T=0$,
and so is $m$, while at $T_{\rm N}$ the degree of nesting has already
been smeared to some extent by thermal effect from the beginning so that
the effect of $h$ on $T_{\rm N}$ is more mild compared to the case of
$m(T=0)$.  The $h$-dependence shown in Fig.\ 2 is exactly
observed in experiments of \urusi~\cite{rf:Mentink,rf:Park}

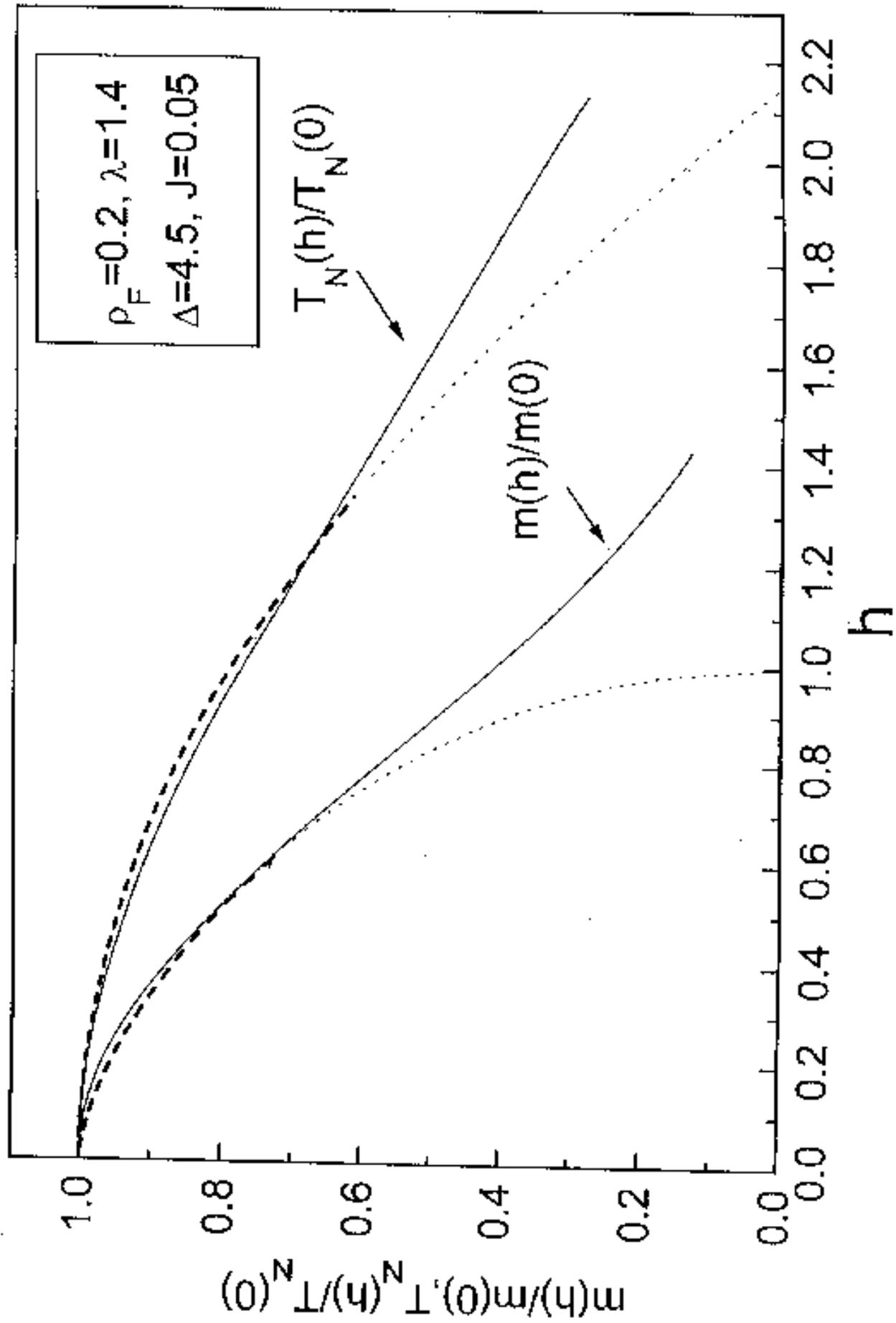
\begin{figure}
\vspace*{7cm}
\caption{The ordered moment m and 
the transition temperature $T_{N}$ vs external magnetic field h. \\
As in Fig.1 we take $c$ as the unit of magnetization and $D$ that of 
energy. The solid lines are our theoretical results for $m(h)$ and 
$T_{N}(h)$, while the dotted and dashed ones are
 phenomenological formulae (\ref{pheno}a,b) with 
$h_{\rm c_{1}}=1.0$ and $h_{\rm c_{2}}=2.15$. 
The dashed ones represent the curves for 
 $m(h)/m(0) > 0.7 $ and $T_{N}(h)/T_{N}(0) > 0.6$ as experiment 
 by Mentink {\it et al}.~\cite{rf:Mentink}
We can see 
good agreement between our results and the phenomenological curves
 for low field region. }
\label{fig:2}
\end{figure}

Experimental results are fit by the following
phenomenological formulae which have different critical
field for magnetic moment and transition temperature.~\cite{rf:Mentink,rf:Park}
\begin{subeqnarray}
m(h) &=& m(0)\sqrt{1-(h/h_{\rm c_{1}})^{3/2}}, \label{pheno} \\
T_{N}(h) &=& T_{N}(0)\bigl[1-\bigl(\frac{h}{h_{\rm c_{2}}}\bigr)^{2}\bigr],
\end{subeqnarray} 
For comparison with our theoretical curves, we show these phenomenological
forms by dotted lines with $h_{\rm c_{1}}=1.0$ 
and $h_{\rm c_{1}}=2.15$.
 One can see good agreement between the phenomenological
formulae and our theoretical results in the low field region. 
These phenomenological formulae are derived below the field at 
$m(h)/m(0)\sim 0.7$ and 
$T_{N}(h)/T_{N}(0)\sim 0.6$,~\cite{rf:Mentink,rf:Park}
 so our results can be regarded to reproduce the
experimental fact that the magnetic moment at 
$T=0$ and the transition temperature appears to have deferent critical  
fields. Mentink {\it et al} have claimed   
the existence of some hidden order parameters from these different 
dependence on magnetic field,  
 we have no need for such an exotic order in our scheme.       
These unconventional features cannot be obtained only from the nesting
property, but are the results of the interplay between the induced-moment
mechanism of ``localized spin" and the nesting property of ``itinerant
fermion" which are well described by the framework of duality model in
a unified way.
\\

\section{Conclusion}
We have shown within the mean-field approximation on the basis of 
itinerant-localized duality model that the small ordered 
moment and large specific 
heat jump observed in URu${}_{2}$Si${}_{2}$     
can be explained by the nesting property of Fermi surface 
with considering the singlet-singlet crystal field scheme. This model 
also explains the apparently different
field dependence of the 
transition temperature and the magnetic moment 
observed by the neutron scattering experiments.~\cite{rf:Mentink}
These ``anomalous'' 
magnetic feature of URu${}_{2}$Si${}_{2}$ 
does not necessarily need the hidden exotic order 
parameter.~\cite{rf:Gorkov} The nesting feature of Fermi surface of 
URu${}_{2}$Si${}_{2}$ is supported by the band 
calculation.~\cite{rf:Ronzing} In addition the induced antiferromagnetism
naturally invokes the orbital order which causes the effects that  cannot 
be understood by simple SDW antiferromagnetism.\\
\indent The realization of 
observed small moment stands in a delicate balance of physical parameters 
to some extent.
We believe  that the values of parameters we take in this papers 
are not so unrealistic and the situation
of real system actually stands in such a balance of physical quantities.
Moreover if we go beyond the mean field level and take into account 
quantum 
fluctuations, we will be able to 
obtain improved results in quantitative level. \\
\indent In this paper we have not taken into account
explicitly  the nature of  the f${}^{2}$-configuration.  
Watanabe and Kuramoto recently pointed out a possibility of 
new kind of metal-insulator transition between the 
localized state f${}^{2}$-configuration with 
crystalline electric field singlet and the itinerant state 
with the Kondo screening.~\cite{rf:Watanabe}
So there may exist the situations of URu${}_{2}$Si${}_{2}$  in which 
we have to consider the nature of 
f${}^{2}$-configuration explicitly as they did. \\
\indent There remains  some other 
important problems for URu${}_{2}$Si${}_{2}$; for 
example, the relation between 
the magnetism and the superconductivity, the properties of quasi-particle 
in the  f${}^{2}$-configuration with singlet  
ground state of crystal field,~\cite{rf:Ikeda} and so on.
These are left  for future investigations.   
\acknowledgement
We are deeply indebted to Professor Y. Kuramoto for valuable 
discussions and critical reading of the manuscripts. 
One of the author (K.M.) has benefited from simplified 
model calculations at very preliminary stage by Kyoko Asano 
and from conversations with N. van Dijk. 
This work is supported by the
Grant-in-Aid for Scientific Research (07640477) and the Grand-in-Aid for
Scientific Research on Priority Areas
``Physics of Strongly Correlated Conductors" (06244104)
from the Ministry of Education, Science, Sports and Culture.


\begin{thebibliography}{99}
\bibitem{rf:Schlabitz}
W. Schlabitz, J. Baumann, B. Pollit, U. Rauchschwalbe, H. M. Mayer,
U. Ahlheim and C. D. Bredl:
Z. Phys. B {\bf 62} (1986) 171.
\bibitem{rf:Fak}
B. F\aa k, C. Vettier, J. Flouquet, F. Bourdarot, S. Raymond,
 A. Verni\`ere, P. Lejay, Ph. Boutrouille, N. R. Bernhoeft, S. T. Bramwell, 
R. A. Fisher and N. E. Phillips:
J. Magn. Magn. Mat. {\bf 154} (1996) 339.
\bibitem{rf:Broholm1}
C. Broholm, J. K. Kjems, W. J. L. Buyers, P. Matthews, T. T. M. Palstra, 
A. A. Menovsky, and J. A. Mydosh:
Phys. Rev. Lett {\bf 58} (1987) 1467.
\bibitem{rf:Palstra}
T. T. M. Palstra, A. A. Menovsky, J. van den Berg, A. J. Dirkmaat, P. H. Kes,
G. J. Nieuwenhuys and J. A. Mydosh: Phys. Rev. Lett. {\bf 55} (1985) 2727.
\bibitem{rf:Maple}
M. B. Maple, J. W. Chen, Y. Dalichaouch, T. Kohara, C. Rossel,
M. S. Torikachvili, M. W. McElfesh and D. J. Thompson: Phys. Rev. Lett.
{\bf 56} (1986) 185.
\bibitem{rf:Mentink}
S. A. M. Mentink, T. E. Mason, S. S\"ullow, G. J. Nieuwenhuys,
A. A. Menovsky, J. A. Mydosh and J. A. A. J.  Peerenboom:
Phys. Rev. B {\bf 53} (1996) R6014.
\bibitem{rf:charge}
P. Santini and G. Amoretti:
Phys. Rev. Lett. {\bf 73}, (1994) 1027 
\bibitem{rf:Nieuwenhuys}
G. J. Niewenhuys: Phys. Rev. B {\bf 35} (1987) 5263.
\bibitem{rf:Radwanski}
R. J. Radwa\'nski: J. Magn. Magn. Mater. {\bf 103} (1992) L1.
\bibitem{rf:Broholm2}
C. Broholm, H. Lin, P. T. Matthews, T. E. Mason, W. J. L. Buyers,
M. F. Collins, A. A. Menovsky, J. A. Mydosh and J. K. Kjems:
Phys. Rev. B {\bf 43} (1991) 12809.
\bibitem{rf:Wang}
Y.-L. Wang and B. R. Cooper: Phys. Rev. {\bf 185} (1969) 696.
\bibitem{rf:Grover}
B. Grover: Phys. Rev. {\bf 140} (1965) A1944.
\bibitem{rf:Sikkema}
A. E. Sikkema, W. J. L. Buyers, I. Affleck and J. Gan:
Phys. Rev. B {\bf 54} (1996) 9322
\bibitem{rf:Kuramoto1}
Y. Kuramoto and K. Miyake:
J. Phys. Soc. Jpn. {\bf 59} (1990) 2831
\bibitem{rf:Kuramoto2}
Y. Kuramoto and K. Miyake:
Prog. Theor. Phys. {\bf 108} (1992) 199.
\bibitem{rf:Okuno}
Y. Okuno, O. Narikiyo and K. Miyake:
J. Phys. Soc. Jpn. {\bf 66} (1997) 2389.
\bibitem{rf:Kohori}
K. Matsuda, Y. Kohori and T Kohara: Physica, B {\bf 230-232} (1997) 351.
\bibitem{rf:Park}
J.-G. Park, K. A. McEwen, S. de Brion, G. Chouteau, H. Amitsuka and
T. Sakakibara: J. Phys. C {\bf 9} (1997) 3065.
\bibitem{rf:Gorkov}
V. Barzykin and L. P. Gor'kov:
Phys. Rev. Lett. {\bf 74} (1995) 4301.
\bibitem{rf:Ronzing}
G. J. Ronzing, P. E. Mijnareds and D. D. Koelling:
Phys. Rev. B. {\bf 43} (1991) 9515.
\bibitem{rf:Watanabe}
S. Watanabe and Y. Kuramoto:
preprint cond-mat/9706134
\bibitem{rf:Ikeda}
H. Ikeda and K. Miyake:
J. Phys. Soc. Jpn. {\bf 66} (1997) 3714.
\end{thebibliography}
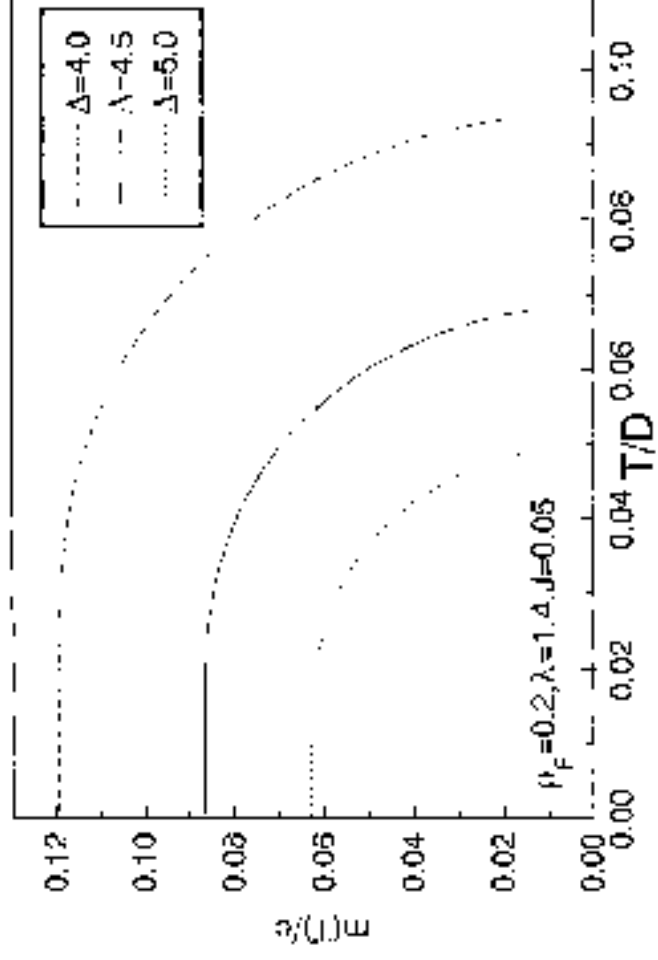
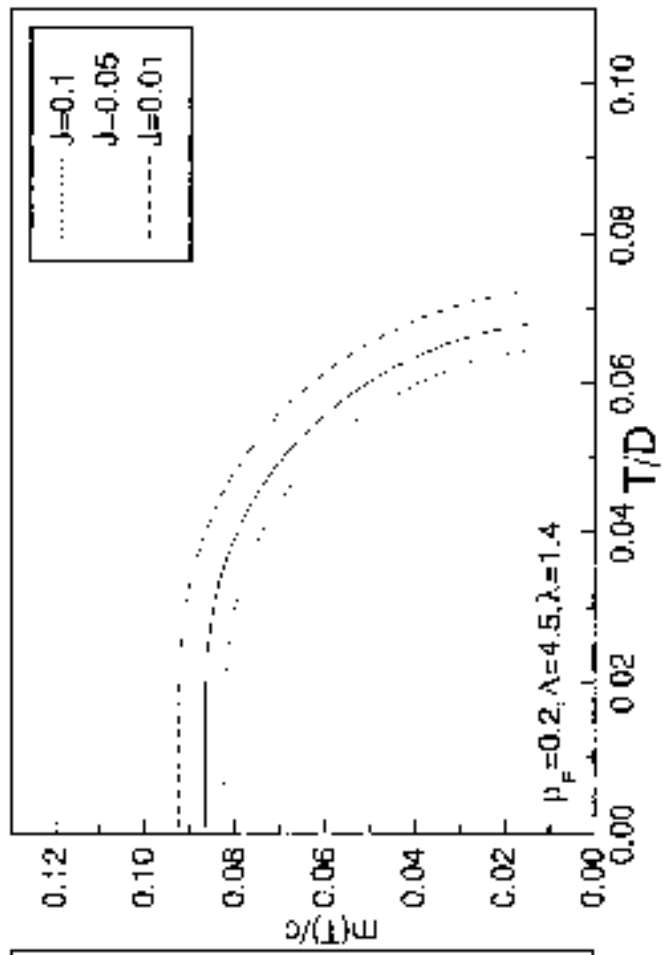
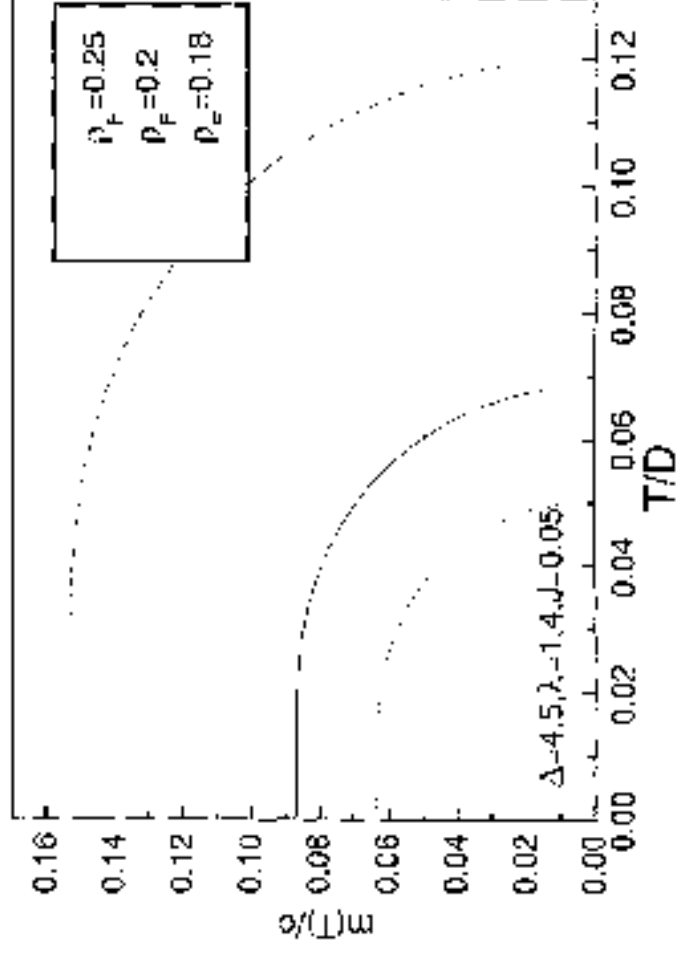
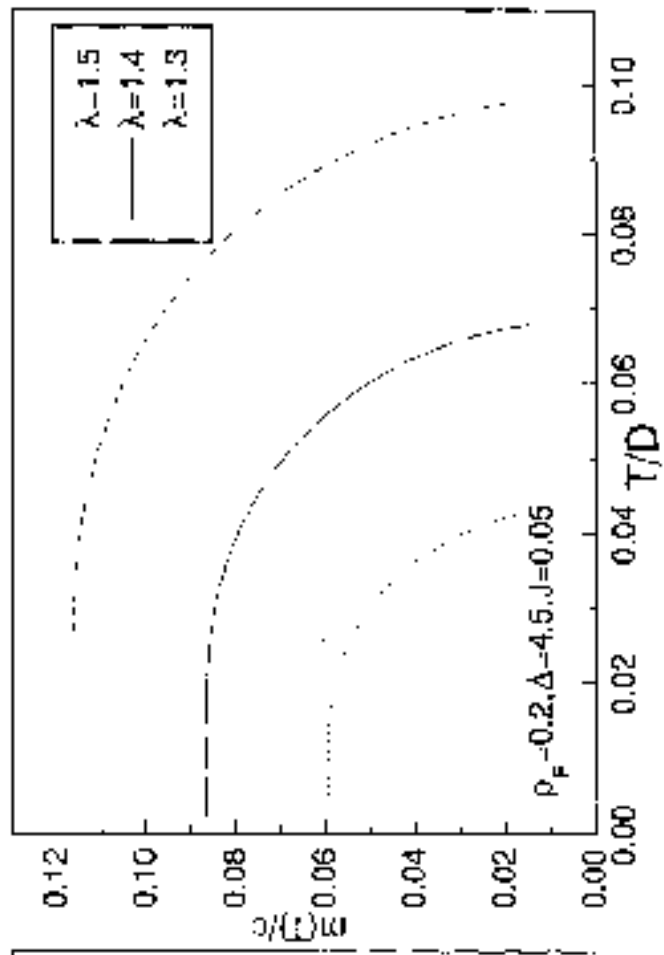
\end{document}